\documentclass[a4paper]{jpconf}

\usepackage{graphicx}
\usepackage{amsmath,amssymb} 
\usepackage{cases}

\newcommand{\bracket}[3]{\langle #1|#2|#3\rangle}
\newcommand{\vecr}{\textbf{r}}

\usepackage[latin1]{inputenc} 
\usepackage[T1]{fontenc}      

\begin{document}
\title{Elastic scattering of a Bose-Einstein condensate at a potential landscape}

\author{Iva B\v rezinov\'a, Joachim Burgd\"orfer}

\address{Institute for Theoretical Physics, Vienna University of Technology,
Wiedner Hauptstra\ss e 8-10/136, 1040 Vienna, Austria, EU}

\author{Axel U. J. Lode, Alexej I. Streltsov, Lorenz S. Cederbaum}

\address{Theoretische Chemie, Physikalisch-Chemisches Institut, Universi\"at Heidelberg, Im Neuenheimer Feld 229, D-69120 Heidelberg, Germany, EU}

\author{Ofir E. Alon}
\address{Department of Physics, University of Haifa at Oranim,
Tivon 36006, Israel}

\author{Lee A. Collins}
\address{Theoretical Division, Los Alamos National Laboratory, Los Alamos, New Mexico 87545, USA}

\author{Barry I. Schneider}
\address{Division of Advanced Cyberinfrastructure, National Science Foundation, Arlington, VA 22230, USA}

\ead{iva.brezinova@tuwien.ac.at}

\begin{abstract}
We investigate the elastic scattering of Bose-Einstein condensates at shallow periodic and disorder potentials. We show that the collective scattering of the macroscopic quantum object couples to internal degrees of freedom of the Bose-Einstein condensate such that the Bose-Einstein condensate gets depleted. As a precursor for the excitation of the Bose-Einstein condensate we observe wave chaos within a mean-field theory.
\end{abstract}

\section{Introduction}
\begin{figure}[t]
\includegraphics[width=16pc]{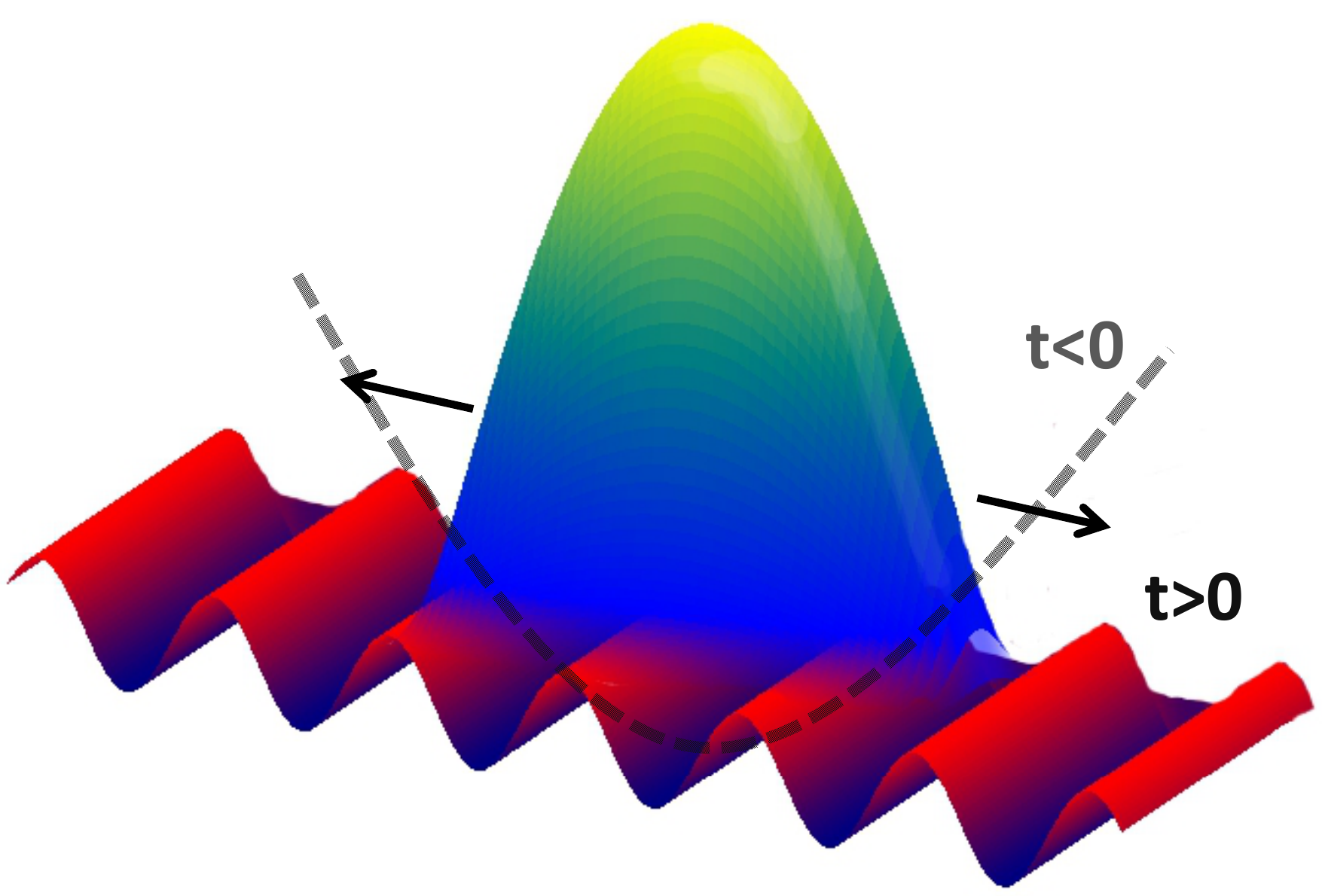}\hspace{2pc}%
\begin{minipage}[b]{16pc}\caption{\label{fig:introduction}A BEC is created in a harmonic trap. At $t=0$ the harmonic trap is switched off, simultaneously a shallow periodic or disorder potential is switched on. We investigate the subsequent non-equilibrium dynamics of the BEC. }
\end{minipage}
\end{figure}

Since the first experimental realization of Bose-Einstein condensates (BECs) in alkali atoms (for a review see e.g.~\cite{proceed98}) a well-controlled macroscopic quantum object is at hand that is governed by the laws of quantum mechanics. A BEC is superfluid, i.e.~it flows without friction below a critical velocity,  
and coherent, i.e.~it can interfere with itself in analogy to laser light. The mechanism underlying superfluidity is the inter-atomic two-body interaction which can be described for ultracold atoms in terms of elastic s-wave scattering. The lowest-lying excitations of this quantum ensemble are collective excitations, i.e.~phonons, the generation of which sets in above a critical threshold, the Landau velocity $v_L$. $v_L$ delimits the velocity range below which the transport proceeds frictionless.\\
Recently, experimental progress has been achieved in investigating BECs in well-controlled external potentials. For example, the properties of BECs in optical lattices created by standing waves of laser light have been studied (for a review see e.g.~\cite{MorObe06}). Another prominent example is the expansion of a BEC in disorder potentials. The goal is to explore for neutral massive particles the mechanism of Anderson localization \cite{And58,BilJosZuoCleSanBouyAsp08,Roa08,DriPolHitHul10}, originally proposed for electrons in condensed matter.\\
In the following we discuss the non-equilibrium dynamics ensuing from the release of an initially harmonics trapped BEC into a static shallow potential landscape. The latter includes either a periodic or a disorder potential (see Fig.~\ref{fig:introduction}). We focus here on the interplay between the collective scattering of the quantum ensemble at the external potential and the two-body scattering within the ensemble. While each of the two is elastic, the coupling from the translational degree of the BEC to the internal degrees of the ensemble can lead to excitation and, eventually, to thermalization of the BEC where no state remains macroscopically occupied. In the presence of such a coupling mediated by the elastic inter-atomic interactions both superfluidity as well as coherence of the BEC may get lost.\\
We describe the dynamics of the system by using both the mean-field approximation based on the Gross-Pitaevskii equation (GPE) as well as the multi-configurational Hartree for bosons (MCTDHB) method which is able to capture excitations out of the condensate state. By comparison between these two theories we are able to delimit the validity of the GPE.\\
One key observation is that the onset of excitations of the BEC becomes evident within the mean-field approximation through the appearance of wave chaos, i.e.~random fluctuations in the solution of the non-linear wave equation, the GPE. A conceptually similar connection between wave chaos and excitations has been found in Ref.~\cite{CasDum97} for a BEC trapped in a harmonic potential with time-dependent frequency. In contrast to our case, energy is not conserved in this system and excitations are created by pumping energy into the system.
%
\section{Theoretical description of Bose-Einstein condensates}
For a theoretical description of a Bose-Einstein condensates the many-body Schr\"odinger equation with the Hamiltonian
\begin{equation}\label{eq:hamiltonian}
H=\sum_{i=1}^{N}\left(-\frac{\hbar^2}{2m}\nabla^2_i+V(\vecr_i)\right)+\frac{1}{2}\sum_{i\neq j=1}^{N}W(\vecr_i-\vecr_j)
\end{equation}
has to be solved. The number of atoms $N$ in the condensate is typically of the order of $N\approx10^4$ or larger such that the Schrödinger equation is not solvable in exact form. Several, approximations, however, can be made which exploit the fact that the BEC is extremely cold and dilute (see below).\\
The external potential $V(\vecr_i)$ in which the initial ground state is prepared is harmonic. At $t=0$ the BEC is released either into a periodic potential or into a disorder potential initiating the non-equilibrium dynamics we are investigating.\\
We choose the initial longitudinal harmonic trap (Fig.~\ref{fig:introduction}) to have a frequency of $\omega_0=5.4$Hz following \cite{BilJosZuoCleSanBouyAsp08} which determines the characteristic length scale $l_0=\sqrt{\hbar/m\omega_0}\approx4.6\mu$m and time scale $t_0=1/\omega_0\approx30$ms of the system. The system is one-dimensional (1D), i.e., radially the BEC is confined by a harmonic trap and is assumed to remain in the ground state with respect to the radial trap through out the whole process. Accordingly, the external 1D potential is of the form
\begin{subnumcases}{V(x)=}
\frac{1}{2}\hbar\omega_0\left(\frac{x}{l_0}\right)^2
&  t<0 \label{eq:harm_pot} \\
V_A \cos{\left (\frac{2\pi}{l}x\right)}\ \text{or}\ V_{\rm disorder}(x) & t>0 \label{eq:exp_pot}.
\end{subnumcases}
Examples of the potentials (Eq.~\ref{eq:exp_pot}) are depicted in the insets of Fig.~\ref{fig:delx}.
\subsection{Interaction potentials}\label{sec:int_pot}
For ultracold and dilute atomic ensembles, the binary interaction potential $W(\vecr_i-\vecr_j)$ can be replaced by a simple pseudopotential which reproduces the inter-atomic scattering in the low energy limit. The interaction potential between two alkali atoms is strongly repulsive for small separations due to the electrostatic repulsion of electrons when the atoms start to overlap, while at large separations attractive van der Waals interactions of induced electric dipole moments dominate giving rise to the typical attractive tail proportional to $-1/r^{6}$ (see e.g.~\cite{PetSmi08}). The interaction potential $W(\vecr)$ is radially symmetric such that during scattering the angular momentum is conserved. For a finite-range potential with range $r_0$ the relative angular-momentum quantum numbers $l$ for which scattering takes place are roughly determined by $l\lesssim kr_0$ where $k$ is the wave number. The range of the van der Waals interactions corresponds approximately to the extent of the least bound state and lies in the range of $3$nm to $10$nm for alkali atoms \cite{Leg08}. For ultracold $^{87}$Rb atoms, e.g., at typical temperatures of $\approx100$nK the thermal de-Broglie wave length is approximately $\lambda_{T}\approx600$nm. Thus, for ultralow-energy scattering $kr_0\ll1$ and we can apply the limit $k\rightarrow0$. In this limit only s-waves scatter and the two-body scattering process is characterized by a single parameter, the s-wave scattering length 
\begin{equation}\label{eq:s_scat_len}
a_s=-\lim_{k\to 0}f(k),
\end{equation}
with $f(k)$ the scattering amplitude. In the case of $^{87}$Rb the scattering length is $a_s\approx5$nm \cite{BurBohEsrGre98} and of the order of $r_0$. When the gas is dilute, i.e.
\begin{equation}
a_s\ll n^{-1/3},
\end{equation}
with $n$ the particle density, the many-body scattering can be reduced to a sequence of independent two-body scattering events and the interaction potential $W(\vecr)$ can be replaced by an effective interaction potential operator (also called the pseudopotential) \cite{Hua87,BloDalZwe08}
\begin{equation}\label{eq:pseudo_pot_cor}
W_{p}(\vecr)=\frac{4\pi\hbar^2a_s}{m}\delta(\vecr)\frac{\partial}{\partial r}r,
\end{equation}
acting on the many-body wavefunction.
The low-energy scattering amplitude resulting from the pseudopotential (Eq.~\ref{eq:pseudo_pot_cor}) is
\begin{equation}\label{eq:scat_amp}
f(k)=\frac{-a_s}{1+ika_s},
\end{equation}
reproducing the correct $k\rightarrow 0$ limit of the original physical potential. \\
In this contribution we consider strongly elongated traps such that the dynamics in radial direction is frozen out (1D traps). We assume the radial trapping to be harmonic with frequency $\omega_r$. To generalize the method of pseudopotentials to 1D we employ a potential which reproduces the two-body low energy scattering in an elongated trap. Using the 3D pseudopotential (Eq.~\ref{eq:pseudo_pot_cor}) Olshanii showed \cite{Ols98} that the 1D pseudopotential is given by
\begin{equation}\label{eq:pseudo_pot_1d}
W(x-x')=2\hbar\omega_ra_s\delta(x-x')=g_{\rm 1D}\delta(x-x'),
\end{equation}
provided $kl_r\ll1$ and $a_s/l_r\ll1$ with $l_r=\sqrt{\hbar/m\omega_r}$. Under these conditions the scattering amplitude can be reduced to that of a 1D delta potential with parameter $g_{\rm 1D}$.
%
%
\subsection{The multiconfigurational time-dependent Hartree for bosons method}\label{sec:mctdhb}
The multiconfigurational time-dependent Hartree for bosons (MCDTHB) method \cite{StrAloCed07,AloStrCed08,StrSakLodAloCed11} is (in principle) an exact many-body theory which allows to describe many-body effects during dynamical evolution of a BEC. The many-body wave function is taken as a superposition of time-dependent permanents
\begin{equation}\label{eq:psi_many}
|\Psi(t)\rangle=\sum_{\{\vec{n}\}}C_{\vec{n}}(t)|\vec{n};t\rangle,
\end{equation}
where $|\vec{n};t\rangle$ corresponds to a state with occupation numbers $\vec{n}=(n_1,...,n_M)$ of the $M$ single-particle orbitals $\{\Phi_k(x,t)\}_{k=1}^{M}$. The sum runs over all sets of occupation numbers $\{\vec{n}\}$ which fulfill $N=\sum_{i=1}^{M}n_i$. In the limit $M\rightarrow\infty$ Eq.~\ref{eq:psi_many} converges to the exact solution of the many-body Schrödinger equation. In practice, only a small number of orbitals can be incorporated for large $N$ as the number of configurations included in Eq.~\ref{eq:psi_many} becomes quickly prohibitively large for large $N$. For example, for $M=3$ and $N=10^4$, Eq.~\ref{eq:psi_many} contains $\approx5\times10^{7}$ configurations.\\
The MCTDHB method efficiently exploits the fact that ultracold atoms may occupy only few orbitals above the condensate state. By dynamically and self-consistently changing the expansion amplitudes $C_{\vec{n}}(t)$ and the orbitals $\Phi_k(x,t)$ even large many-body systems can be treated accurately. The equations of motion correspond to a set of coupled linear Schr\"odinger equations for $C_{\vec{n}}(t)$ and coupled non-linear partial differential equations for the orbitals $\Phi_k(x,t)$ (see e.g.~\cite{AloStrCed08}). (For details on the present numerical implementation see \cite{BreLodStrAloCedBur12}).
\subsection{The Gross-Pitaevskii equation}\label{sec:gpe}
The GPE can be viewed as the limiting case of the MCTDHB method with one single orbital, i.e.~$M=1$, describing the condensate wavefunction only. Accordingly, the many-body wavefunction reduces to a product state of the form
\begin{equation}\label{eq:mb_prod_ansatz}
\Psi(x_1,...,x_N,t)=\prod_{i=1}^{N}\psi(x_i,t).
\end{equation}
In the absence of two-body interaction, Eq.~\ref{eq:mb_prod_ansatz} is an exact solution of the independent-particle Schrödinger equation for the condensate. It obviously neglects two-particle correlations for non-zero two-particle interactions and can account for interactions only on the mean-field level. The product state Eq.~\ref{eq:mb_prod_ansatz} is assumed to be a good approximation in the dilute regime where
\begin{equation}\label{eq:dilute_cond}
a_s\ll n^{-1/3},
\end{equation}
(see e.g.~\cite{Leg08,LieSeiSolYng06}). The time-dependent GPE can be derived from the variational principle
\begin{equation}\label{eq:func_der_gpe}
\frac{\delta \bracket{\Psi}{H-i\hbar\frac{\partial}{\partial t}}{\Psi}}{\delta \psi^*(x,t)}=0
\end{equation}
with $H$ from Eq.~\ref{eq:hamiltonian} (in the present case in its 1D form) and the product ansatz (Eq.~\ref{eq:mb_prod_ansatz}) for $\psi$ resulting in
\begin{eqnarray}\label{eq:gpe_1d}%
i\hbar\frac{\partial \psi(x,t)}{\partial t}=-\frac{\hbar^2}{2m}\frac{\partial^2 \psi(x,t)}{\partial x^2}+V(x)\psi(x,t)+g_{1D}N|\psi(x,t)|^2\psi(x,t),
\end{eqnarray}
first derived in \cite{Pit61,Gro61}. For the present parameters (see Ref.~\cite{BreColLudSchBur11,BreLodStrAloCedBur12} for details) the strength of the non-linearity in the GPE is given by $g_{1D}N\approx390\hbar\omega_0l_0$.\\
We note that the use of the exact scattering length of the two-body interaction within the interaction term of the GPE (Eq.~\ref{eq:gpe_1d}) has been questioned (see e.g.~\cite{Gel09}). As an alternative, the scattering length derived from the (first) Born approximation has been suggested, in particular for the attractive branch of the interaction potential. This approximation employs the fact that $\psi(\vecr)$ varies only weakly in the scale given by the large potential. Accordingly, the Hartree-like interaction term $\int d^3r \int d^3r' |\psi(\vecr)|^2|\psi(\vecr')|^2W(\vecr-\vecr')$  can be replaced by $\int d^3r|\psi(\vecr)|^4\int d^3r'W(\vecr')$ \cite{Fri06}. The term $\int d^3r' W(\vecr')$ is then proportional to the $k\rightarrow0$ limit of the scattering amplitude in Born approximation
\begin{equation}
\int d^3 r' W(\vecr')=\lim_{k\to 0}\left(-\frac{4\pi\hbar^2}{2\bar m}f^B\right)=\frac{4\pi\hbar^2}{2\bar m}a_s^B,
\end{equation}
where the superscript $B$ denotes the Born approximation. In turn one obtains the same prefactor as in Eq.~\ref{eq:pseudo_pot_cor} but with $a_s$ replaced by $a_s^B$. In the following we employ the more established strategy by using the exact scattering length $a_s$ rather than $a_s^{\rm B}$. The underlying reasoning is that the applicability of the Born approximation requires the potential to be weak (see e.g.~\cite{BasDal02}) a condition not generally satisfied. Moreover,
for attractive potentials, $a_s^B$ does not account for the change of sign (from positive to negative) at the transition to a new bound state and thus does not account for the effect of Feshbach resonances already observed in ultracold atoms (see e.g.~\cite{ChiGriJulTie10}).
%
\subsection{Observables}\label{sec:obs}
The simplest and most important benchmark observable for a comparison with the experiment as well as between the mean-field and the many-body dynamics is the single-particle density. Within the MCTDHB method the density is given by 
\begin{eqnarray}\label{eq:density}
\rho(x,t)&=&N\int dx_2...dx_N\;\Psi^*(x,x_2,...,x_N,t)
\Psi(x,x_2,...,x_N,t)\nonumber \\
&=&\sum_{m,n=1}^M\rho_{m,n}(t)\Phi_m^*(x,t)\Phi_n(x,t),
\end{eqnarray}
where the elements $\rho_{m,n}(t)$ of the one-particle density matrix are given in terms of a combination of the amplitudes $C_{\vec{n}}(t)$ and the corresponding occupation numbers contained in $\vec{n}$ (see Ref.~\cite{AloStrCed08}). Upon diagonalization of Eq.~\ref{eq:density} the density in terms of the natural orbitals $\Phi_i^{\rm NO}(x,t)$ and their occupation numbers $n_i^{\textrm{NO}}(t)$ is obtained as
\begin{equation}\label{eq:den_NO}
\rho(x,t)=\sum_{i=1}^{M}n_i^{\rm{NO}}(t)|\Phi_i^{\rm{NO}}(x,t)|^2.
\end{equation}
The presence of a BEC is signalled by the ``macroscopic'' (of order $N$) occupation of one single state \cite{PenOns56}. In the case of a fragmented BEC several natural orbitals are occupied macroscopically. In the following we denote the condensate state as $\Phi_{i=1}^{\textrm{NO}}(x,t)$ and its occupation as $n_{i=1}^{\textrm{NO}}(t)$. All other states $\Phi_i^{\textrm{NO}}(x,t)$ with $i>1$ are referred to as excited states.
\\
A sensitive probe for correlations within the many-body system and for the breakdown of the GPE is given by the normalized two-particle correlation function \cite{HanTwi56,Gla07,SakStrAloCed08} 
\begin{equation}\label{eq:g2}
g^{(2)}(x_1',x_2',x_1,x_2,t) \equiv \frac{\rho^{(2)}(x_1',x_2',x_1,
x_2,t)}
{\sqrt{{\rho(x_1,t)}{\rho(x_2,t)}{\rho(x'_1,t)}{\rho(x'_2,t)}}}
\end{equation}
which is closely related to the second-order coherence of quantum optics.
In $g^{(2)}$ the reduced two-body density matrix
\begin{eqnarray}\label{eq:rho_2}
\rho^{(2)}(x_1',x_2',x_1,x_2,t)=
N(N-1)\int\; dx_3...dx_N\Psi^*(x_1',x_2',x_3,...,x_N,t) \Psi(x_1,x_2,x_3,...,x_N,t)
\end{eqnarray}
enters. Within the GPE the reduced two-body density matrix is a product of one-body wave functions (compare with Eq.~\ref{eq:mb_prod_ansatz}). Thus, $|g^{(2)}|=1$ for all times, i.e., full second-order coherence is a generic feature of the GPE. Conversely, deviations from $|g^{(2)}|=1$ indicate many-body corrections. For a finite number of particles, the correlation function for a fully condensed system (all particles in the condensate state) is given by $|g^{(2)}|=1-1/N$ and reaches unity for $N\rightarrow\infty$. Deviations from these limiting values measure (anti)correlation in the system and admixture to the many-body wavefunction beyond a single particle product state.
%
\section{Scattering at periodic and disorder potentials}
We consider propagation of a BEC in shallow periodic and disorder potentials (Eq.~\ref{eq:exp_pot}). The disorder potential is constructed numerically by equidistantly spaced Gaussians of width $\sigma$ with random amplitude \cite{BreColLudSchBur11}. The potential is normalized to $V_A=\sqrt{\langle V^2(x)\rangle}$. A shallow potential implies that $V_A\ll e$, where $e$ is the mean energy per particle which is closely related to the chemical potential $\mu$. In the examples presented in the following we use $V_A=0.2e$. For the periodic potential we choose $l=5.8\xi$ with $\xi=\hbar/\sqrt{4m\mu}$ being the healing length. For the disorder potential we chose $\sigma=0.7\xi$. Both potentials are depicted in the inset of Fig.~\ref{fig:delx}.
\subsection{Wave chaos within the mean-field theory}
%
\begin{figure}[t]
\begin{minipage}{18pc}
\includegraphics[width=18pc]{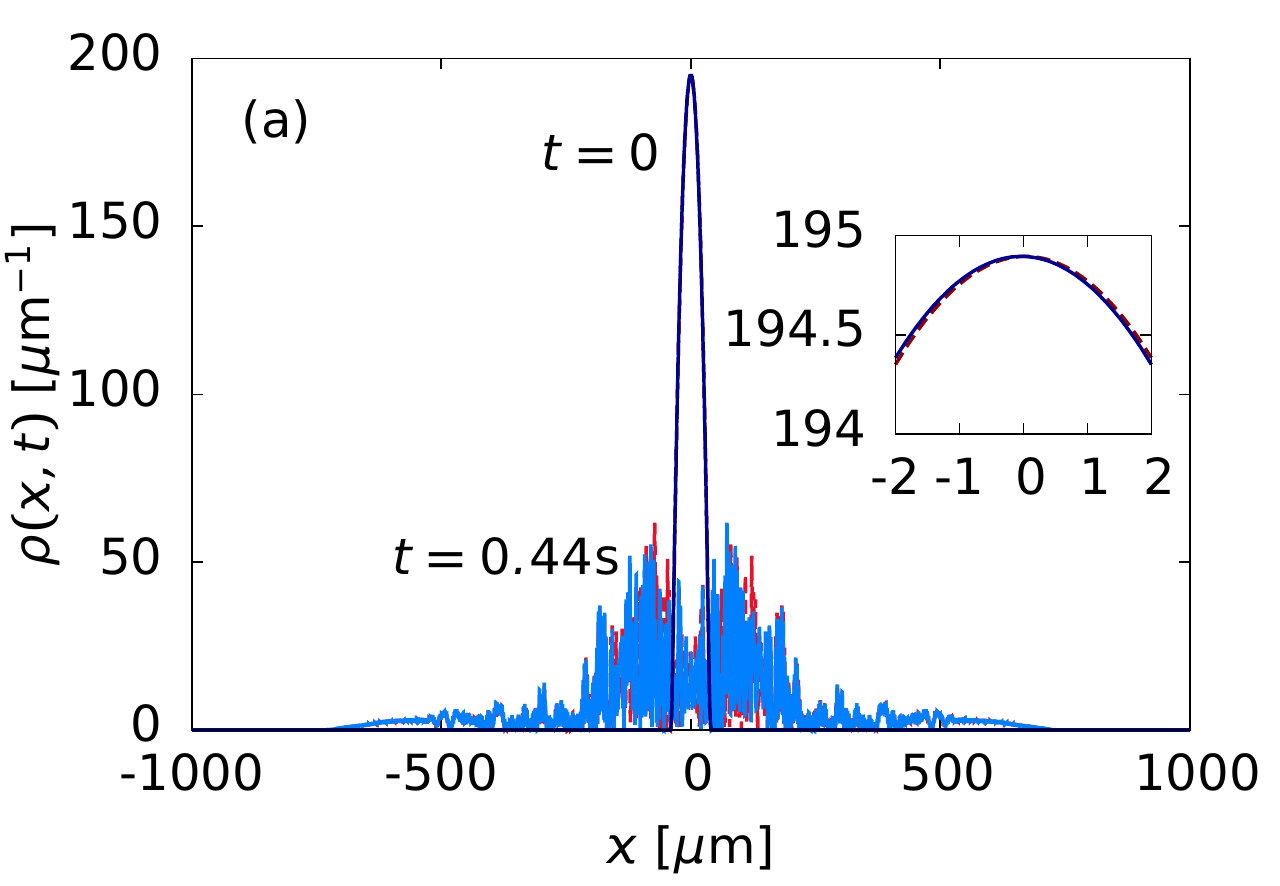}
\end{minipage}\hspace{2.pc}%
\begin{minipage}{18pc}
\includegraphics[width=18pc]{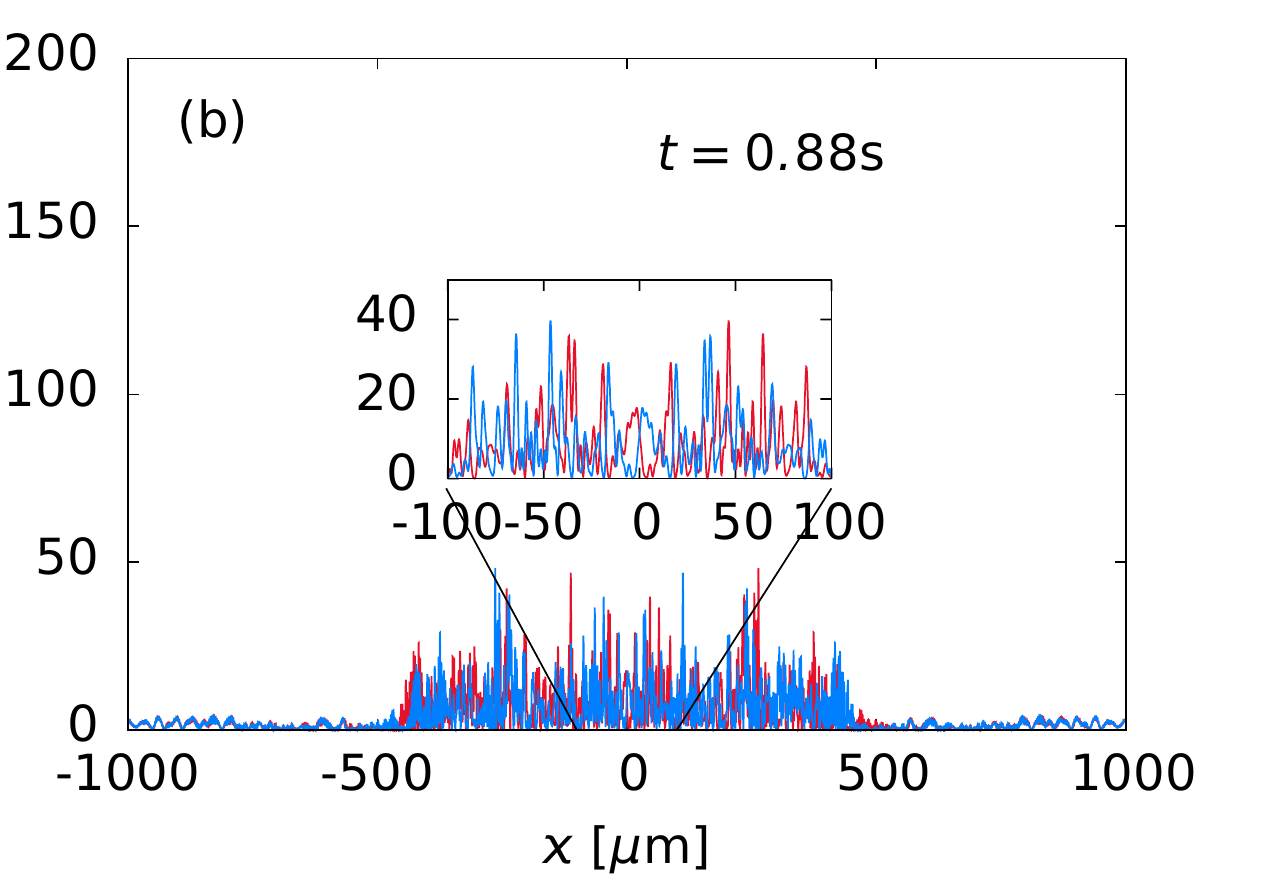}
\end{minipage} 
\caption{Snapshots of particle densities at three instants of time during time evolution within the GPE in a periodic potential with $l\approx5.6\xi$ and $V_A=0.2e$. Starting point are two slightly different initial densities (a) created using $\psi_{1,2}(x)\propto(1\pm\alpha x)\psi_{G}(x)$ with $\alpha=10^{-4}$, where $\psi_{G}(x)$ is the ground state of the initial harmonic trap, see Fig.~\ref{fig:introduction} (for details see also Ref.~\cite{BreColLudSchBur11}). During time propagation, (a) and (b), large discrepancies emerge.}\label{fig:wave_chaos}
\end{figure}
We solve the GPE (Eq.~\ref{eq:gpe_1d}) numerically for the scenario described in Fig.~\ref{fig:introduction}. We first calculate the ground state of the harmonic trap $\psi_G(x)$ [see Fig.~\ref{fig:wave_chaos} (a)] and propagate it in the presence of the shallow periodic or disorder potential and in the absence of the harmonic potential. Since the GPE is a non-linear differential equation, its solutions behave distinctly different from the solutions of the linear Schr\"odinger equation. We have found that the solutions of the GPE exhibit wave chaos for the systems discussed in the present contribution. Wave chaos is conceptually analogue to classical chaos where initially close points in phase space exponentially separate as a function of time. In the case of wave chaos the exponential separation takes place in Hilbert space. In Fig.~\ref{fig:wave_chaos} the ``separation'' of two initially close wavefunctions is depicted. The ``separation'' of wavefunctions manifests itself by the emergence of ``random'' fluctuations at the length scale of the external potential. The separation is, indeed, exponential and can be quantified by a generalized Lyapunov exponent \cite{BreColLudSchBur11}. We interpret the emergence of wave chaos within the mean-field theory as an indication that the overall elastic process of a BEC scattering at an external potential is associated with excitations of internal decrees of freedom of the condensate.
%
%
\subsection{Collective excitations}
%
\begin{figure}[t]
\includegraphics[width=20pc]{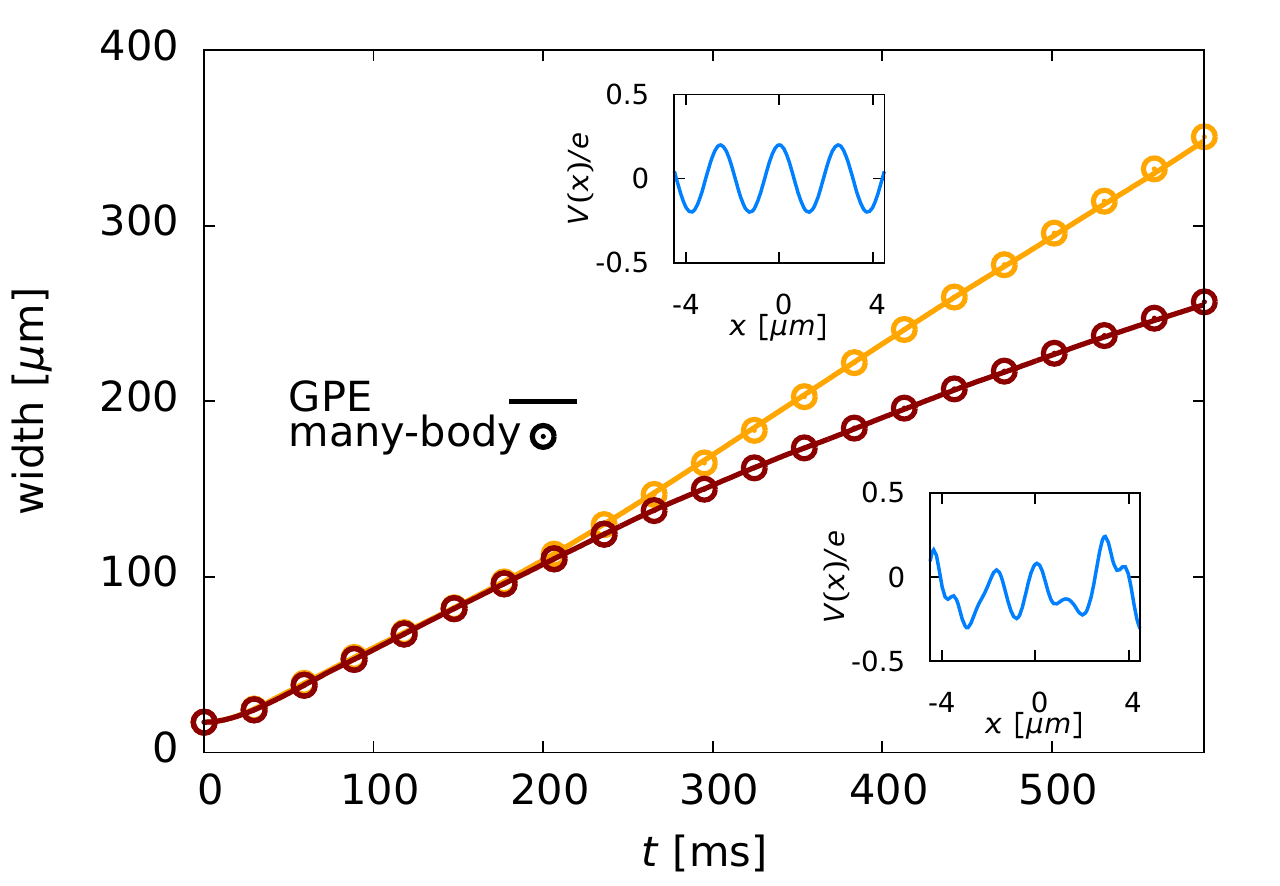}\hspace{0pc}%
\begin{minipage}[b]{14pc}\caption{\label{fig:delx} The width $\Delta x$ of the atomic cloud as a function of time as given by the GPE and the MCTDHB. The upper (orange) curve corresponds to the periodic potential (upper inset), the lower (dark red) curve corresponds to the disorder potential (lower inset).}
\end{minipage}
\end{figure}
\begin{figure}[t]
\includegraphics[width=20pc]{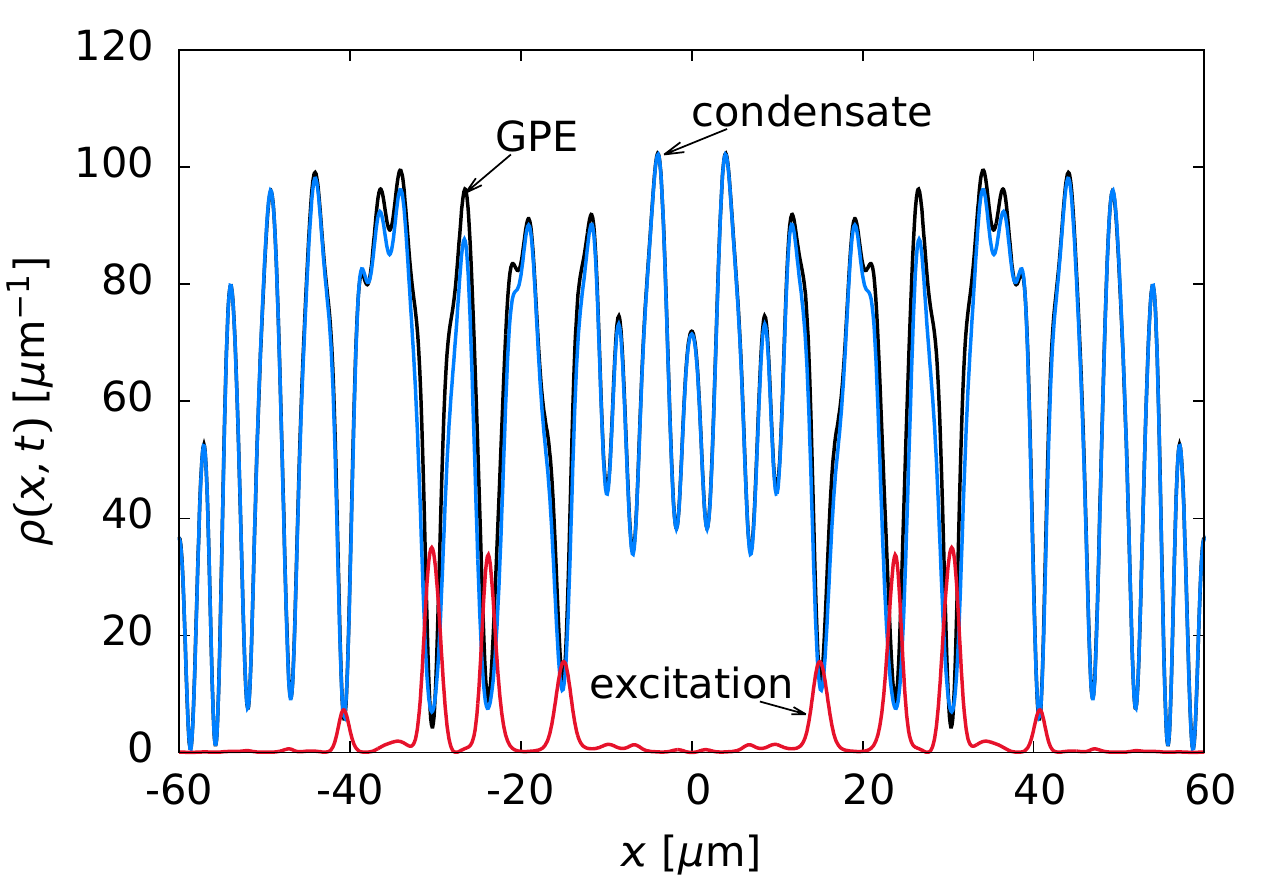}\hspace{0pc}%
\begin{minipage}[b]{14pc}\caption{\label{fig:depletion} Particle density as determined by the GPE and the MCTDHB for $N=10^4$ particles at $t\approx120$ms. Within the MCTDHB the condensate density is given by $n_1^{\rm NO}|\Phi_{1}^{\rm NO}(x,t)|^2$, the excitations are given by $n_2^{\rm NO}|\Phi_{2}^{\rm NO}(x,t)|^2$ (see Eq.~\ref{eq:den_NO}). About $\approx5\%$ of the particles have left the condensate.}
\end{minipage}
\end{figure}
\begin{figure}[t]
\begin{minipage}{22pc}
\includegraphics[width=22pc]{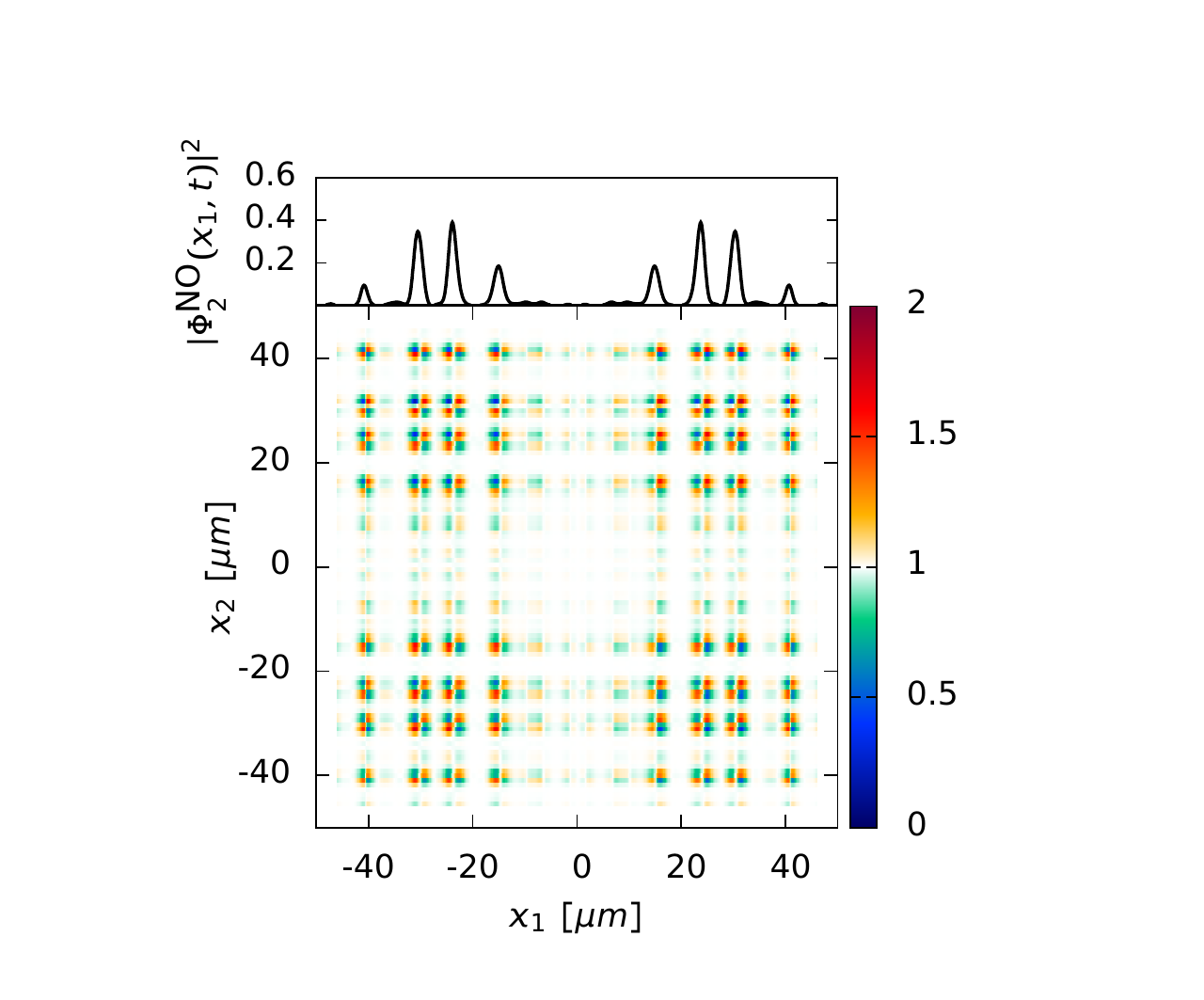}
\end{minipage}\hspace{-3.pc}%
\begin{minipage}{22pc}
\includegraphics[width=22pc]{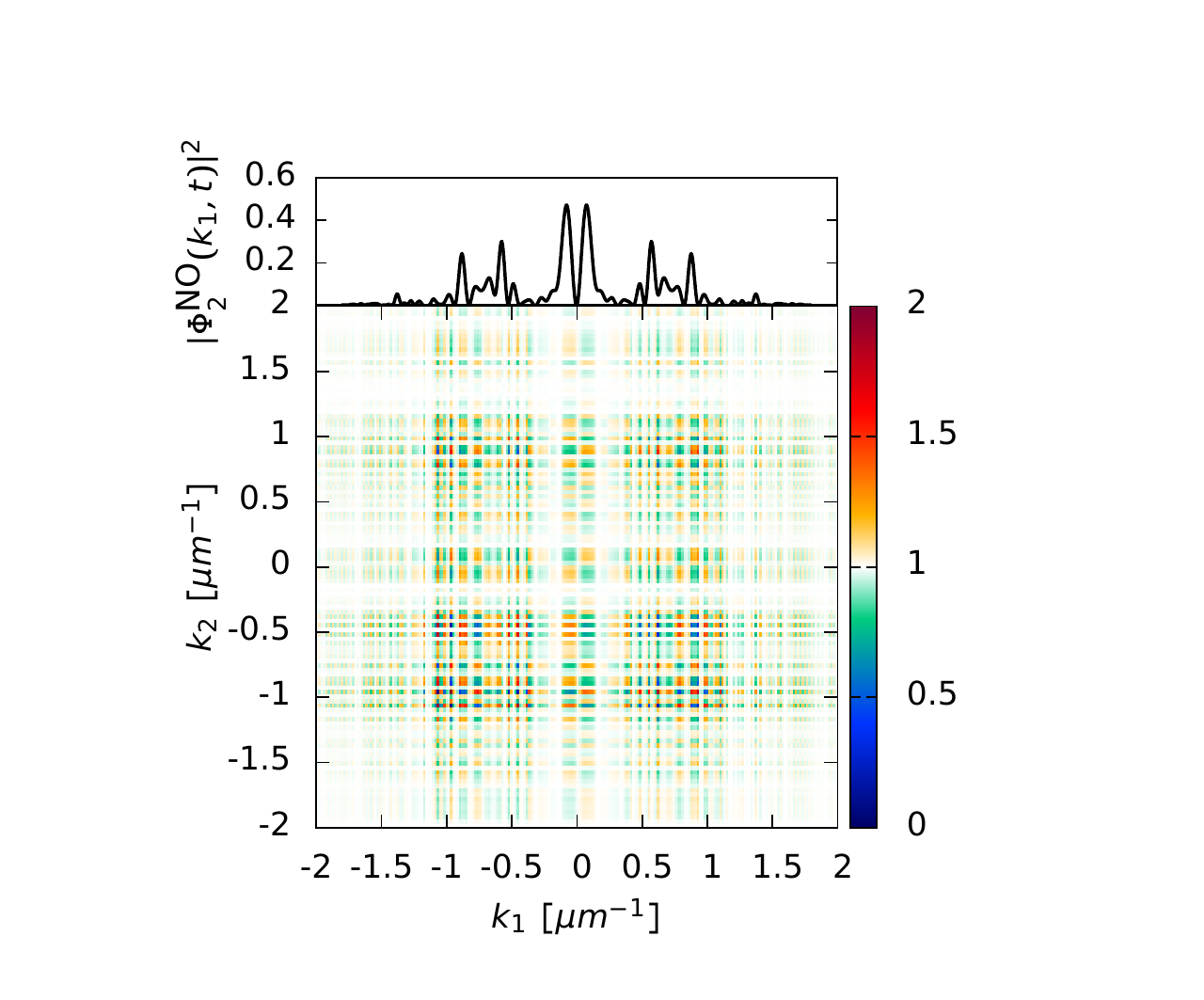}
\end{minipage}
\caption{The diagonal element of the $g^{(2)}$-function (Eq.~\ref{eq:g2}) in coordinate space (left) and in momentum space, i.e.~Fourier transform of $g^{(2)}(x_1',x_2',x_1,x_2;t)$, (right) at $t\approx120$ms. In the upper part of each figure the natural orbital corresponding to excitations is depicted.}\label{fig:g2}
\end{figure}
To demonstrate the relation between wave chaos within the GPE and collective excitations we use the MCTDHB method. We observe that the divergence of initially nearby wavefunctions and the concurrent development of random yet deterministic fluctuations as seen in Fig.~\ref{fig:wave_chaos} is associated with the depletion of the ground-state orbital (Fig.~\ref{fig:depletion}). The time scale for the emergence of wave chaos within the GPE on one hand, and the occupation of excited states within the MCTDHB method on the other hand, nearly coincide \cite{BreLodStrAloCedBur12} (in the present case $t\approx120$ms). In the present example with $N=10^4$ atoms only $M=2$ orbitals are included within MCTDHB, for numerical reasons. Thus all excitations are represented by one orbital, $\Phi_{2}(x,t)$. This is, of course, a crude approximation, nevertheless, on a qualitative level, it allows to estimate both the time scale as well as the rate with which excitations are created \cite{BreLodStrAloCedBur12}. We find that the collective elastic scattering of the BEC at a potential landscape leads to the excitation of the initially fully coherent quantum ensemble.\\
At $t\approx600$ms the occupation of the excited state corresponds to $n_2\approx50\%$ indicating a destruction or at least a strong fragmentation of the condensate. Clearly, the GPE is not capable to accurately describe this internal excitations dynamics. It is now of considerable interest to inquire into the validity of the GPE for coarse-grained one-body observables in the regime where excitations are prevalent. As an example we consider the width of the atomic cloud, $\Delta x=\sqrt{\langle x^2\rangle-\langle x\rangle^2}$, where $\langle x\rangle$ denotes the expectation value for position. Fig.~\ref{fig:delx} displays $\Delta x$ for propagation in both the periodic and the disorder potential. While $\Delta x$ grows linearly in time for the periodic potential showing ballistic behavior, the expansion in the disorder potential obeys a power law $\Delta x\propto t^a$ with $a<1$ and decreasing with time (Fig.~\ref{fig:delx}). For long times we observe subdiffusive spreading with $a<0.5$ in qualitative agreement with \cite{LapBodKriSkoFla10}, see Ref.~\cite{BreColLudSchBur11}.\\
It is now remarkable that the GPE and the MCTDHB method almost perfectly agree for $\Delta x$ for both the periodic and the disorder potential (Fig.~\ref{fig:delx}). We thus arrive at the conclusion that the validity of the GPE extends well into the regime where excitations out of the condensate become important, provided that one is interested in coarse-grained one-body observables.\\
Conversely, probing and quantifying the failure of the GPE requires observables that are more sensitive to correlations. We suggest to measure the second-order coherence function $g^{(2)}$ (Eq.~\ref{eq:g2}).
While a pure finite-number condensate with $n_1=100\%$ is (almost) perfectly coherent, excitations reduce coherence and lead to deviations of $g^{(2)}$ from one (Fig.~\ref{fig:g2}). We show the $g^{(2)}$-function in both coordinate and momentum space. In both cases $g^{(2)}$ can be measured via density-density correlations either in situ or after time-of-flight (see e.g.~\cite{ManBucBet10}). Remarkably, the pattern of the $g^{(2)}$-function closely follows that of the excited orbital (upper panel of Fig.~\ref{fig:g2}).\\

\section{Summary and Conclusions}
In this contribution we have discussed BECs expanding in shallow one-dimensional potential landscapes. We have shown that the collective scattering of the initially fully coherent quantum ensemble at the potential landscape leads to the excitation of internal degrees of freedom of the many-body system. On a microscopic scale, the coupling of the translational decrees of freedom to internal degrees of freedom is mediated by inter-atomic elastic s-wave scattering. While the mean-field description in terms of the Gross-Pitaevskii equation cannot directly account for the excitations out of the condensate we have found that the non-linearity of the GPE induces wave chaos whose appearance is closely correlated with the excitations. It can be viewed as the indicator on the mean-field level of the excitation dynamics of the underlying many-body system. By comparing to a many-body theory, the MCTDHB method, we showed that the random yet deterministic fluctuations are associated with the occupation of excited natural orbitals and the depletion of the condensate. 
This process goes hand in hand with the reduction of coherence of the BEC and thus becomes accessible experimentally through the $g^{(2)}$ function. Remarkably, the GPE remains predictive for averaged observables such as the width of the expanding atomic cloud, even though it is not able to capture excitations out of the condensate state.
\section*{Acknowledgments}
This work has been supported by the FWF program ``CoQuS'' and the FWF-SFB 041 ``ViCoM''. Calculations have been performed on the Vienna Scientific Cluster.
\section*{References}
\bibliographystyle{iopart-num}
\providecommand{\newblock}{}

\end{document}